\documentclass[journal=jpccck,manuscript=article]{achemso}

\usepackage{chemformula} 
\usepackage[T1]{fontenc} 
\usepackage[labelfont=bf]{caption}
\usepackage{comment}

\author{Jia Meng}
\affiliation[University of Alberta]
{Department of Chemical and Materials Engineering, University of Alberta, Edmonton, AB, T6G 1H9, Canada}
\author{Jae Bem You}
\email{jaebem@ualberta.ca}
\affiliation[University of Alberta]
{Department of Chemical and Materials Engineering, University of Alberta, Edmonton, AB, T6G 1H9, Canada}
\alsoaffiliation[University of Twente]
{Physics of Fluids Group, Max Planck Center Twente for Complex Fluid Dynamics, JM Burger Center for Fluid Dynamics, Mesa+, Department of Science and Technology, University of Twente, Enschede 7522 NB, The Netherlands}
\author{Xuehua Zhang}
\email{xuehua.zhang@ualberta.ca}
\affiliation[University of Alberta]
{Department of Chemical and Materials Engineering, University of Alberta, Edmonton, AB, T6G 1H9, Canada}
\alsoaffiliation[University of Twente]
{Physics of Fluids Group, Max Planck Center Twente for Complex Fluid Dynamics, JM Burger Center for Fluid Dynamics, Mesa+, Department of Science and Technology, University of Twente, Enschede 7522 NB, The Netherlands}

\title[An \textsf{achemso} demo]
{Viscosity-mediated Growth and Coalescence of Surface Nanodroplets}

\abbreviations{}

\begin{document}


\begin{abstract}

Solvent exchange is a simple method to produce surface nanodroplets on a substrate for a wide range of applications by displacing a solution of good solvent, poor solvent and oil (Solution A) by a poor solvent (Solution B). In this work, we show that the growth and coalescence of nanodroplets on a homogeneous surface is mediated by the viscosity of the solvent. 
We show that at high flow rates of viscous Solution B, the final droplet volume deviates from the scaling law that correlates final droplet volume to the flow rate of non-viscous Solution B, reported in previous work. We attribute this deviation to a two-regime growth in viscous Solution B, where transition from an initial, fast regime to a final slow regime influenced by the flow rate. Moreover, viscous solution B hinders the coalescence of growing droplets, leading to a distinct bimodal distribution of droplet size with stable nanodroplets, in contrast to a continuous size distribution of droplets in non-viscous case. We demonstrate that the group of small droplets produced in high viscosity environment may be applied for enhanced fluorescence detection with higher sensitivity and shorter response time. The finding of this work can potentially be applied for mediating the size distribution of surface nanodroplets on homogeneous surface without templates.

\end{abstract}

\section{Introduction}
Surface nanodroplets have attracted significant research interest, due to various applications for micropatterning, \cite{yang2020} optical lenses,\cite{Xu2020acsami,qian2019} chemical and biological analysis,\cite{garcia2017,zeng2016analchem} and crystallization \cite{Stephens2011}. Tuning the size distribution of droplets is important for droplet-based sensing,\cite{wang2018pressuresensor} biochemical assays,\cite{dittrich2019droplet} miniaturized substrates for cell culture\cite{levkin2018droplet}, fabrication of nanostructures from ionic liquids, \cite{yu2020} and many others. 
Currently methods such as inkjet-based droplet printing,\cite{Yu2015Dropletarray,Cole2017PNAS} and droplet deposition on pre-patterned substrate\cite{zhang2004uniform,Paulssen2018micrpattern} are widely used to for producing droplet with a defined size. However,  
complicated procedures, dedicated equipment, or long process time are often required for the formation of large quantity of nanodroplets. Some of the methods are not applicable for producing surface droplets immersed in a liquid medium.

Recently, solvent exchange process has been developed as a bottom-up approach to reliable formation of surface nanodroplets, providing control over final droplet volume, composition as well as number density.\cite{Zhang9253,lohseRMP,qian2019} These droplets are usually up to a few nanometers in the maximal apex and several to tens microns in lateral diameter. 
The solvent exchange process is versatile, thus enabling novel applications of surface nanodroplets including liquid-liquid extraction \cite{miaosi2019small,lohse2016towards} and 
\textit{in-situ} determination of partition coefficient \cite{miaosi2019analchem}. In a typical solvent exchange process, surface nanodroplets are formed by exposing the substrate surface to a good solvent (Solution A) which is subsequently displaced by a poor solvent (Solution B). At the mixing front of Solution A and B, droplets nucleate and grow due to local oversaturation of droplet liquid in the mixture. At the end of the process, the final droplet volume can be controlled by tuning several parameters such as flow rate, compositions of Solution A and B as well as the channel dimensions.\cite{zhangCR2015} Changing flow rate of Solution B can vary droplet sizes without changing the solution composition for solvent exchange. Droplet size control by the flow rate can be advantageous when solubility phase diagram of the ternary mixture in Solution A is not available. For a given channel dimension and oil concentration, the droplet volume scales with the flow rate following the scaling law $Vol_{f} \sim Pe^\frac{3}{4}$. Here, $Vol_{f}$ is the final droplet volume per unit area and $Pe$ is the P{\'e}clet number defined as $Pe = Q/(wD)$ where $Q$ is the flow rate of Solution B, $w$ is the channel width and $D$ is the diffusivity of oil.\cite{Zhang9253} 

The base size, spacing and position of the droplets by solvent exchange can be tuned precisely on a surface with chemical micropatterns \cite{bao2015small}.  However, on a homogeneous  smooth surface the size distribution of droplets is uncontrolled, due to stochastic nature of nucleation, growth and coalescence from oversaturation in the surrounding \cite{Beysens2006,Park2016,xu2017collective,yang2019}. In a similar process of droplet formation from condensation \cite{knobler1986,zhangAM2007,zhangCR2015}, the viscosity of the solution cast on the surface has been shown to an important parameter to eliminate the undesirable coalescence of droplets. When two droplets are surrounded by a highly viscous fluid, their coalescence is delayed due to slower displacement of liquid between them \cite{chiesa2006investigation,luo2019electrocoalescence}. 

 In this work, we study the effect of solution viscosity on the droplet formation by solvent exchange. 
 In particular, we demonstrate that when viscosity of Solution B is higher than that of Solution A, the final droplet size becomes independent of flow rate at sufficiently high flow rates. 
Importantly, the high viscosity of solutions surrounding the droplets hinders their coalescence, resulting in a bimodal size distribution. We will demonstrate that the small droplets produced via hindering coalescence enhance the fluorescence intensity in detection with higher sensitivity and shorter time.  The findings of this work may provide a method for producing surface nanodroplets as templates for materials fabrication with hierarchical structures, such as polymer films with bimodal pore sizes \cite{yasuga2018self,zhang2015breathfigure,kaneko2019} or microlens with largely different focusing length.




\section{Experimental}
\subsection{Substrate and solutions}
\indent A Si substrate was rendered hydrophobic by coating a monolayer of octadecyltrichlorosilane (OTS) (95\%,  Alfa Aesar) following the protocol reported elsewhere.\cite{Zhang9253} The substrate of OTS coated Si was sonicated in ethanol (90\%, Fisher Scientific) and dried by air before use.
\\
\indent The solvent exchange process to generate surface nanodroplets was performed in a custom-built fluid chamber with an inlet and an outlet (Fig. S1a). The chamber was 5.8 cm long, 1.3 cm wide and 0.97 mm high. Glass was used as the top cover, allowing for observation of the droplets using an upright optical microscope. Solution A consisting of ethanol, water and 1,6-hexanediol diacrylate (HDODA) (99\%, Alfa Aesar) was injected into the chamber using a syringe pump (NE-1000, Pumpsystems Inc.). Subsequently, a solution of glycerol (99.5\%, Fisher BioReagents) in HDODA-saturated water (Solution B) was injected into the chamber to displace the Solution A and form surface nanodroplets. The viscosity of Solution B is adjusted by the concentration of glycerol.\cite{cheng2008formula,volk2018density,Grunberg1949MixtureLF} To minimize the influence of density difference between glycerol and water, the solvent exchange process was carried out in a vertical position such that both solutions were introduced from bottom to top (Fig. \ref{diagram}a and Fig. S1b). Composition of Solution A and B were varied to test the effect on the droplet formation including groups 1, 2 and 3 with different viscosity of Solution B and groups 4, 5 and 6 with different oil concentrations.  Viscosity of Solution A and Solution B are shown in Table 1.
\subsection{Characterization of the droplets}
Optical microscope (Nikon ECLIPSE Ni) equipped with a camera (Nikon DS-Fi3) was used to obtain images of the droplets. The droplet volume per unit surface area (V/A) and size distribution were obtained by analyzing more than 1 mm$^2$ from each image using ImageJ. From each image, the occupied surface area of each droplet was obtained and its corresponding lateral radius ($R$) and volume ($V$) could be estimated using the surface area and contact angle ($\theta$).
\\
\indent For the measurement of contact angle of the nanodroplets, surface profile of polymerized HDODA droplets were obtained via atomic force microscope (AFM) by tapping mode. A representative AFM image of droplet is attached in shown in Fig. 1b. The polymerized nanodroplets were obtained by following the method reported in our previous work.\cite{Zhang9253} Contact angle was then calculated by fitting a circle equation to the obtained profile. The contact angle of the droplets on OTS-Si were measured to be $\sim 9^{\circ}$ with 7.4 cP solution B and $\sim 8^{\circ}$ with 12.8 cP and 18.0 cP solution B. The contact angle is assumed to be independent of droplet size, following observation from a previously reported work.\cite{Zhang9253}

\subsection{Extraction of Nile Red from water using nanodroplets}
Extraction of Nile Red (Sigma Aldrich, $\lambda_{ex} \sim$ 554 nm, $\lambda_{em} \sim$ 638 nm) from water was performed using the droplets produced from the solvent exchange process with high viscosity Solution B. Here, we used HDODA as the oil component, making Solution A at ethanol : water : HDODA $=$ 48.1 : 48.1 : 3.8 (in wt\%). We used glycerol to increase the viscosity of Solution B to $\mu_B \sim$ 7.4 cP. 
Droplets were formed on OTS-Si via standard solvent exchange with Solution A and B at flow rate of 96 mL/h. Subsequently, a sample solution containing trace of Nile Red in water was injected at the same flow rate. As the sample solution was introduced into the chamber, Nile Red was extracted to the HDODA droplets, and fluorescence signal was detected in the droplets under green laser. The fluorescence signal intensity of the Nile Red in droplets was measured using self-written MATLAB codes.

\section{Results and discussion}
\subsection{Effect of viscosity on droplet growth dynamics}
\begin{figure}[ht]
\centering
\includegraphics[scale=1]{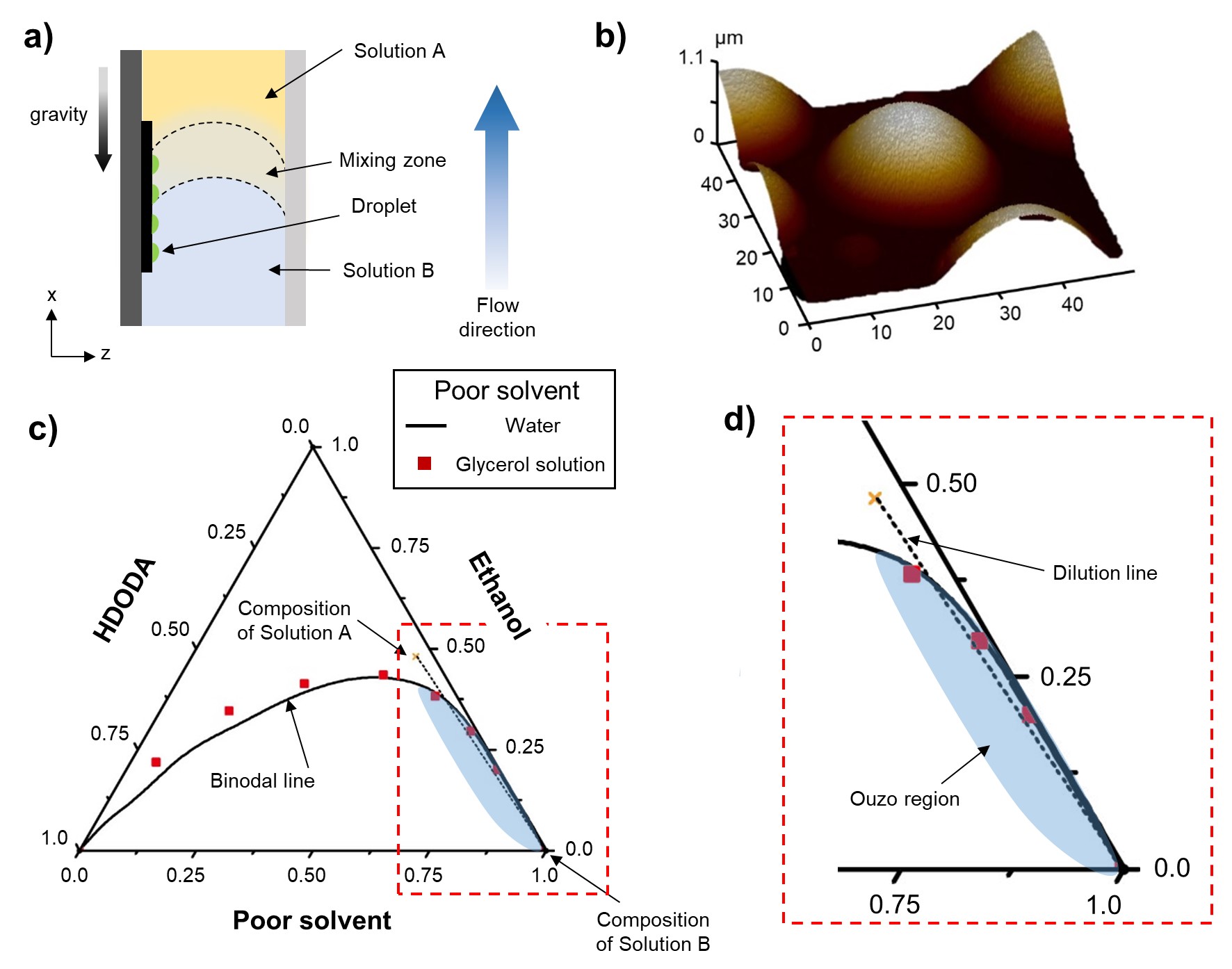}
\captionsetup{font={small}}
\caption{a) Schematic of solvent exchange process. Since the flow direction is from bottom to top, in positive x direction, any influence of gravity on the flow profile can be prevented. b) Atomic force microscope (AFM) image of a polymerized droplet. c) Ternary phase diagram of HDODA, ethanol and water plotted along with that of glycerol solution. Addition of glycerol does not change the phase diagram much, especially near Ouzo region shown in color blue. During solvent exchange, Solution B (either water or glycerol solution) gradually displaces Solution A following the dilution path (dashed line). As the composition crosses the binodal line, phase separation occurs forming droplets. d) The zoom-in view of the phase diagram near Ouzo region.}
\label{diagram}
\end{figure}

\begin{figure}[ht]
\centering
\includegraphics[scale=1]{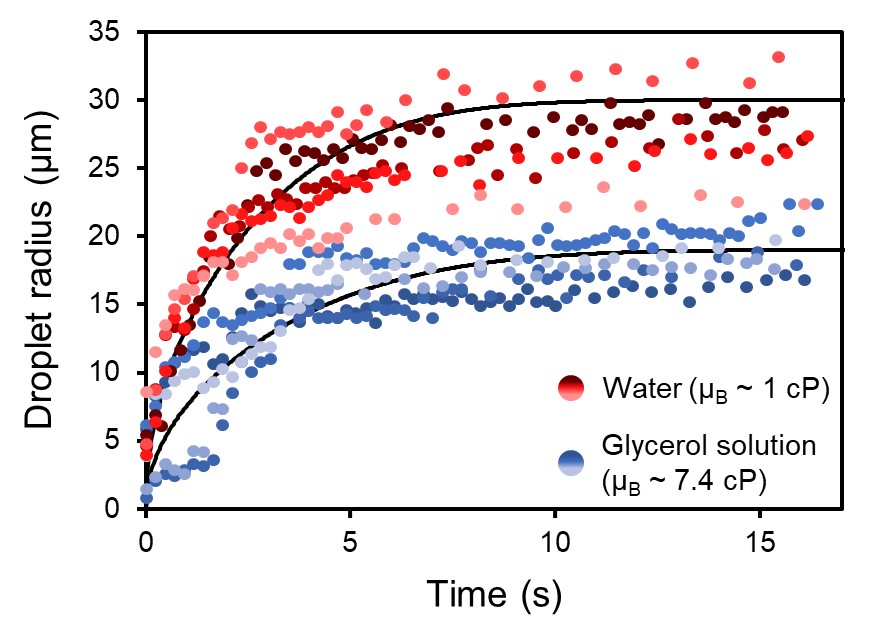}
\captionsetup{font={small}}
\caption{Droplet growth dynamics for the cases with water (red) and glycerol solution (blue) as Solution B. For the case of glycerol solution, the droplet radius plateaus after $\sim$3 s. In contrast, droplet continues to grow for the case of water.}
\label{dynamics}
\end{figure}

The solvent exchange process relies on the phase separation caused by the oversaturation of oil as the mixture of oil/good solvent/poor solvent (Solution A) is replaced by the oil-saturated poor solvent (Solution B). As shown in the ethanol/water/HDODA ternary phase diagram (Fig.\ref{diagram}c),  Solution A is a homogeneous solution with the compositions above the binodal line. As Solution B is introduced into the chamber to gradually displace Solution A, diffusive mixing between the two solutions takes place and the composition of the mixture cross the Ouzo region surrounded by spinodal and binodal lines.\cite{vitale2003,Zemb2015} The change in composition triggers liquid-liquid phase separation due to local oversaturation in the mixing zone and finally leads to the formation of droplets on the substrate. Surface nanodroplets are also produced even if the viscosity of Solution B is adjusted using glycerol. The phase diagram obtained with glycerol solution remains similar to that of water, especially near Ouzo region (Fig. \ref{diagram}d). This similarity suggests that liquid-liquid phase separation and droplet formation would also occur using viscous Solution B in solvent exchange.
\\
\indent However, when the viscosity of Solution B is increased by addition of glycerol, droplet growth dynamics is clearly different from the non-viscous case. To compare the growth dynamics between non-viscous ($\mu_{B} \sim$ 1 cP) and viscous cases ($\mu_{B} \sim$ 7.4 cP), we track the lateral radius ($R$) of droplets for a period of more than 15 s (Fig. \ref{dynamics}). As shown in Fig. \ref{dynamics}, at a given time, droplets grow faster in the case of using water as Solution B compared to the case of using glycerol solution. Although droplets grow at a similar rate initially for the first 3 s, the droplet size reaches a plateau to a value between 15 $\sim$ 20 $\mu$m after $\sim$3 s in the case of glycerol solution. On the other hand, the droplet size for water case lies between 20 $\sim$ 30 $\mu$m and still continues to grow with time, with growth dynamics following $R(t) \sim C_{0}erf(t/\tau)^\frac{1}{2}$ where $C_{0}$ is a pre-factor, and $\tau$ is growth time, as reported in our recent work. \cite{dyett2018growth} The values used for fitting were $C_{0} =$ 30 and $\tau \sim$ 5.66 for non-viscous case of water and $C_{0} =$ 19 and $\tau \sim$ 0.29 for the viscous case of glycerol solution. 
The fact that both of these values are different for non-viscous and viscous cases indicates substantial difference in their growth dynamics.
\\
\indent It is worth mentioning that in this set of our experiments, a horizontal setup had to be employed to track the change in droplet size with time during the solvent exchange. In such configuration, gravity may influence the growth of the droplets.\cite{yu2015gravitational}. To determine the influence of gravity, Archimedes number comparing the gravitational force to viscous force was calculated for both cases. 

$$Ar = \frac{gh^3\Delta{\rho}}{\nu^2\rho_{B}} \eqno{(3)} $$

where g $=$ 9.81 $m/s^2$ is the gravitational acceleration, $h$ is the channel height (250 $\mu$m), $\Delta{\rho}$ is the density difference between Solution A ($\rho_{A}$) and Solution B ($\rho_{B}$), and $\nu$ is the kinematic viscosity of Solution B at composition of glycerol:water $=$ 51:49. In the non-viscous case, Ar $\sim$ 16 whereas in the viscous case Ar $\sim$ 0.7 indicating that gravity would influence the growth in the non-viscous case, leading to smaller droplet volume as reported in literature.\cite{yu2015gravitational} Even though Ar $>$ 1 in the non-viscous case, the droplet growth is still much faster compared to the viscous case clearly demonstrating the influence of viscosity on the dynamics of droplet growth.

\subsection{Effect of viscosity on final droplet size}
\begin{figure}[ht]
\centering
\includegraphics[scale=1.0]{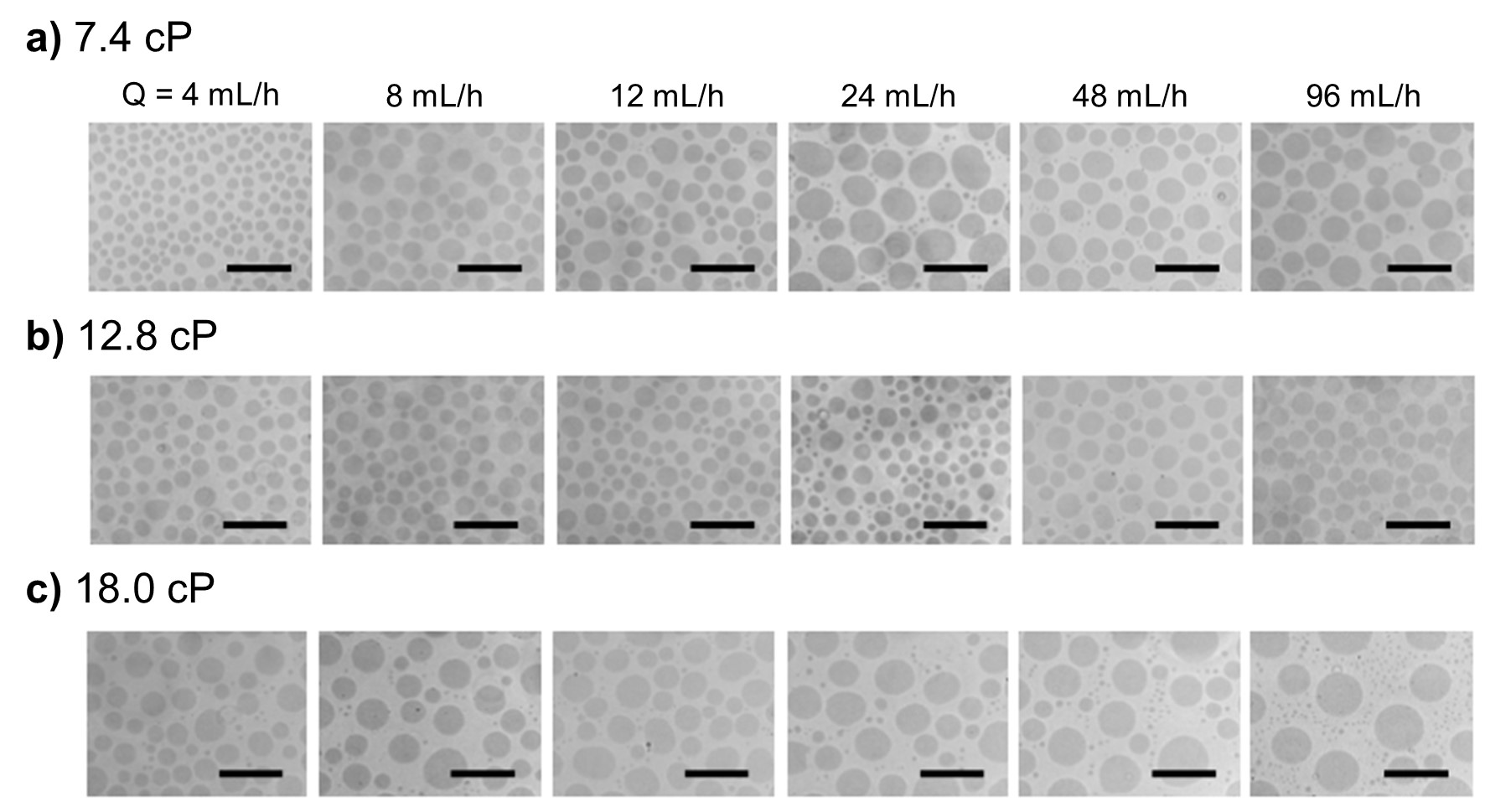}  
\captionsetup{font={small}}
\caption{Optical images of droplets formed at flow rates from 4 to 96 mL/h with different viscosity of Solution B. a) $\mu_B \sim$ 7.4 cP, b) $\mu_B \sim$ 12.8 cP, c) $\mu_B \sim$ 18.0 cP. Scale bar: 50 $\mu m$.}
\label{snapshot}
\end{figure}

\begin{figure}[ht]
\centering
\includegraphics[scale=1]{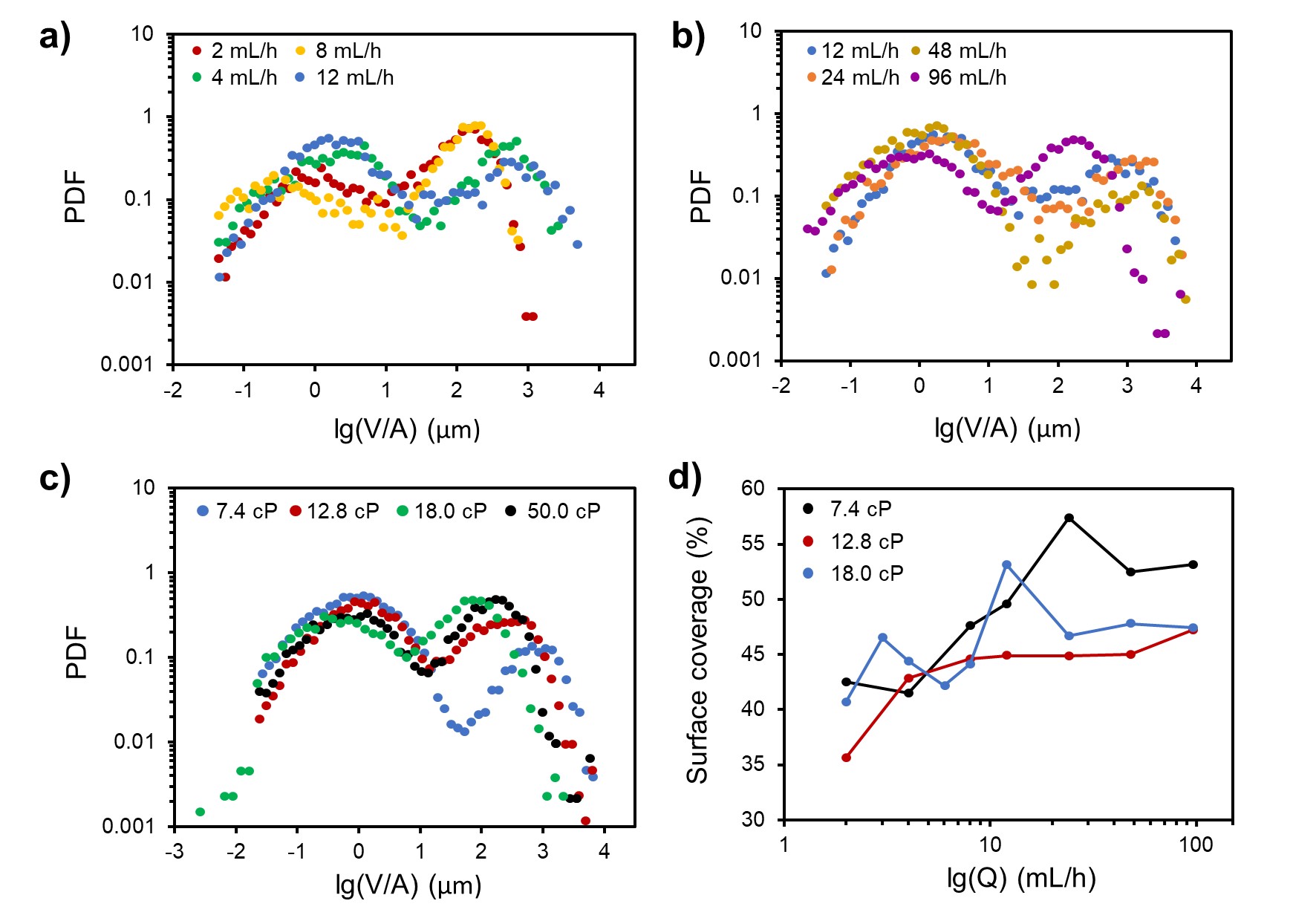}
\captionsetup{font={small}}
\caption{Probability distribution function (PDF) of droplet volume per unit area (V/A) produced (in $\mu m$) with Solution B viscosity of 18.0 cP at a) 2 to 12 mL/h and b) 12 to 96 mL/h. c) PDF of V/A produced at 96 mL/h for Solution B viscosity from 7.4 to 50 cP. d) Surface coverage versus flow rate with different solution B viscosity. The surface first increases with flow rate, while plateau after the flow rate reaching 12 mL/h.} 
\label{PDF}
\end{figure}

\indent Fig. \ref{snapshot} shows the optical images of the surface nanodroplets formed by Solution B with viscosity of 7.4 cP to 18.0 cP at flow rate from 4 mL/h to 96 mL/h. For all values of viscosity, the size of droplets was observed to increase with increase in the flow rate. However, above a certain flow rate, no significant change in size was observed.  Instead smaller droplets were formed in between the already-growing large droplets. The trend was more pronounced at higher viscosity (i.e. 18.0 cP). Notably, for 18.0 cP, small droplets were observed starting at 4 mL/h with no drastic increase in size of already formed droplets at higher flow rates. However, for the case with viscosity $\sim$ 7.6 cP, small droplets were formed at a higher flow rate and the size of primary droplets were observed to increase with flow rate until it plateaued at Q $=$ 24 mL/h.
\\
\indent Fig. \ref{PDF}a and b show the probability distribution function (PDF) of the volume per unit surface area (V/A) of droplets formed by Solution B with viscosity $\sim$ 18.0 cP for different flow rates. Here, the viscosity of Solution A was kept at $\sim$ 1 cP. For all cases, a bimodal distribution is observed with a peak located at lg(V/A) $\sim$ 0.5 corresponding to the small droplets and another peak near lg(V/A) $\sim$ 2.5 to 3.0 corresponding to large droplets. This is clearly different from the case when both solutions are non-viscous ($\mu \sim$ 1.0 cP), where size distribution was continuous with one broad peak indicating a continuous size distribution.\cite{Zhang9253} 
\\
\indent The bimodal distribution of droplet size is also evident for other viscosity values. Fig. 4c shows the PDF of droplet sizes produced at 96 mL/h for Solution B with viscosity from 7.4 to 50 cP. All the distributions have two distinct peaks, reflecting the presence of two groups of droplets with vastly different sizes. Notably, while the sizes of large droplets are noticeably different for different viscosity values, small droplets are similar in size regardless of viscosity. Fig. \ref{PDF}d shows the surface coverage measured at different flow rates for the Solution B viscosity of 7.4, 12.8 and 18.0 cP. When the flow rates are low, surface coverage increases with flow rate. However, after a certain point (i.e. 12 mL/h), the surface coverage fluctuates between 45 and 55\%, consistent with the distribution of droplet sizes shown in Fig. \ref{PDF}a,b.


The other notable difference between viscous and nonviscous cases is the influence of flow rate on the droplet size. High viscosity of Solution B leads to the independence of droplet size on flow rate,  reflected as deviation from the scaling law of $Pe^{3/4}$, previously confirmed for the non-viscous solution B.\cite{Zhang9253} Fig. 5a shows the lg(V/A) as function of lg(Q) with dotted lines representing best fit lines with 3/4 slope according to the scaling law. In all cases, the data deviates from the power of 3/4 after certain lg(Q) value but remains at a plateau. As the oil concentration is varied, the plateau of the droplet size is still observed, that is, the final droplet volumes becomes independent of the flow rate when Solution B is fast enough (Fig. \ref{plateau}). 

\subsection{Flow rate-independence and two-regime growth of droplets at high viscosity }

The universality of the scaling law correlating the droplet volume per unit surface area to the $Pe$ with the power of {3/4} has been verified through several studies using different variables such as chamber height, solvent exchange direction, oil with different viscosity, and initial solution composition\cite{qian2019}. It is surprising that the higher viscosity of Solution B compared to that of Solution A leads to flow-rate independence, especially in the high flow rate regime. We propose the following two-regime growth model to rationalize the effect of viscosity on the dynamics of droplet growth during the solvent exchange. 

When using viscous Solution B, the entire process of the droplet growth may be divided into a fast-growth regime (regime 1) and a slow-growth regime (regime 2) as sketched in Fig. \ref{regime}. In the fast-growth regime, the viscosity in the mixing zone is still low due to the higher amount of non-viscous Solution A compared to Solution B, and droplet growth follows the process same as the non-viscous Solution B, as reported before \cite{Zhang9253}. Briefly,  $ \dot{m}=4 \pi \rho_{oil} R^2 \dot{R}=4\pi D R^2 \partial_r c |_R$, where $m$ is droplet mass, $\rho_{oil}$ is density of oil, $R$ is lateral droplet radius, $D$ is diffusivity and $c$ is concentration of oil. Here, the concentration gradient at the interface ($\partial_r c |_R$) is the gradient between the oil concentration in the bulk flow ($c_\infty$) and at the interface of droplets ($c_{s,poor}$) divided by the thickness of the concentration boundary layer ($\lambda_1$).\cite{Zhang9253} In the mixing zone, the diffusion of oil into the droplets is influenced by Taylor dispersion,\cite{aris1956dispersion,taylor1953dispersion,grossmann2004fluctuations} so  $\lambda_1 \sim R/\sqrt{Pe}$. Therefore, the concentration gradient becomes:
$$\partial_r c |_R \sim \frac{c_\infty(t)-c_{s,poor}}{\lambda_1} \sim c_{s,poor} \frac{\zeta(t)}{\lambda_1} \sim c_{s,poor} \sqrt{Pe(t)}R^{-1} \zeta (t) \eqno{(4)}$$

where $\zeta(t)$ denotes the oversaturation profile of oil. Then, using the concentration gradient and mass balance around the droplet, we obtain: 
$$ R \dot{R} \sim \frac{D(t) c_{s,poor}}{\rho_{oil}}\sqrt{Pe(t)} \zeta (t) \eqno{(5)}$$
Here, both $D$ and $Pe$ change with time during the droplet formation as the viscosity of the medium in the mixing zone increases. In regime 1, because the ratio of Solution B compared to Solution A is still low, the change of $D$ and $Pe$ with time is not significant, therefore the droplet growth is enhanced by increase in $Pe$ of Solution B. In non-viscous case, both $D$ and $Pe$ are almost constant during the solvent exchange, the above theoretical analysis explains the scaling law of final droplet size with $Pe^{3/4}$ as reported in the previous work. \cite{Zhang9253} 
\\
\indent The high viscosity of Solution B eventually starts to influence the mass transport for droplet growth. The momentum boundary layer ($\delta$) above the solid surface increases with time by following the Blasius solution ($\delta \sim 1/ \sqrt{Re}$). 
Here the local diffusivity $D$ and Reynolds number ($Re=\frac{\rho Ud}{\mu}$ with $\rho$ density of fluid, $U$ local velocity, and $d$ characteristic length) at a fixed point on the surface decreases with time as the ratio of Solution B in the mixing zone increases. As $\delta$ becomes large enough, the concentration boundary layer is nested in the momentum boundary layer.  The influence from the momentum boundary layer on droplet growth becomes negligible at $\lambda_1 \sim R/\sqrt{Pe} \sim R$ (i.e the concentration boundary layer thickness is comparable to the droplet size). The growth of droplets now transitions into the second regime (regime 2) where the concentration gradient around the growing droplet is: 

$$\partial_r c |_R \sim c_{s,poor} \frac{\zeta{(t)}}{\lambda_2} \sim c_{s,poor} R^{-1} \zeta{(t)}\eqno{(6)}$$

 $$ R \dot{R} \sim \frac{D(t) c_{s,poor}}{\rho_{oil}}\zeta{(t)}\eqno{(7)} $$

Different from eq. 5, $Pe$ is not included in eq. 7. The final droplet volume $R_f$ is the combination of both regimes:

$$\int_{0}^{R_f}Rdr \sim \frac{ c_{s,poor}}{\rho_{oil}}[\int_{0}^{\tau_1}D(t) \sqrt{Pe(t)}\zeta(t) dt + \int_{\tau_1}^{\tau_2}D(t) \zeta(t) dt]\eqno{(8)}$$

Here $\tau_1$ is time duration of regime 1, and $\tau_2$ is time duration of regime 2. The total growth time $\tau = \tau_1+\tau_2$, influenced by the channel height, diffusivity and Taylor-Aris dispersion. \cite{dyett2018growth}  For two extreme cases, the droplet growth either follows $Pe$ scaling law $Vol_f \sim Pe^{3/4} \sim Q^{3/4}$, same as the case of non-viscous Solution B\cite{Zhang9253}, or is independent of the flow rate or thus $Pe$. We are not able to provide a solution for eq. (8), due to complicated inter-correlation of various parameters. However, the above analysis shows that as a result of more viscous Solution B than Solution A, the final droplet size is influenced by flow rate $Q$ (or $Pe$), growth time in each regime ($\tau_1$ and $\tau_2$), diffusivity and their dependence with time.  

\begin{figure}[ht]
\centering
\includegraphics[scale=1]{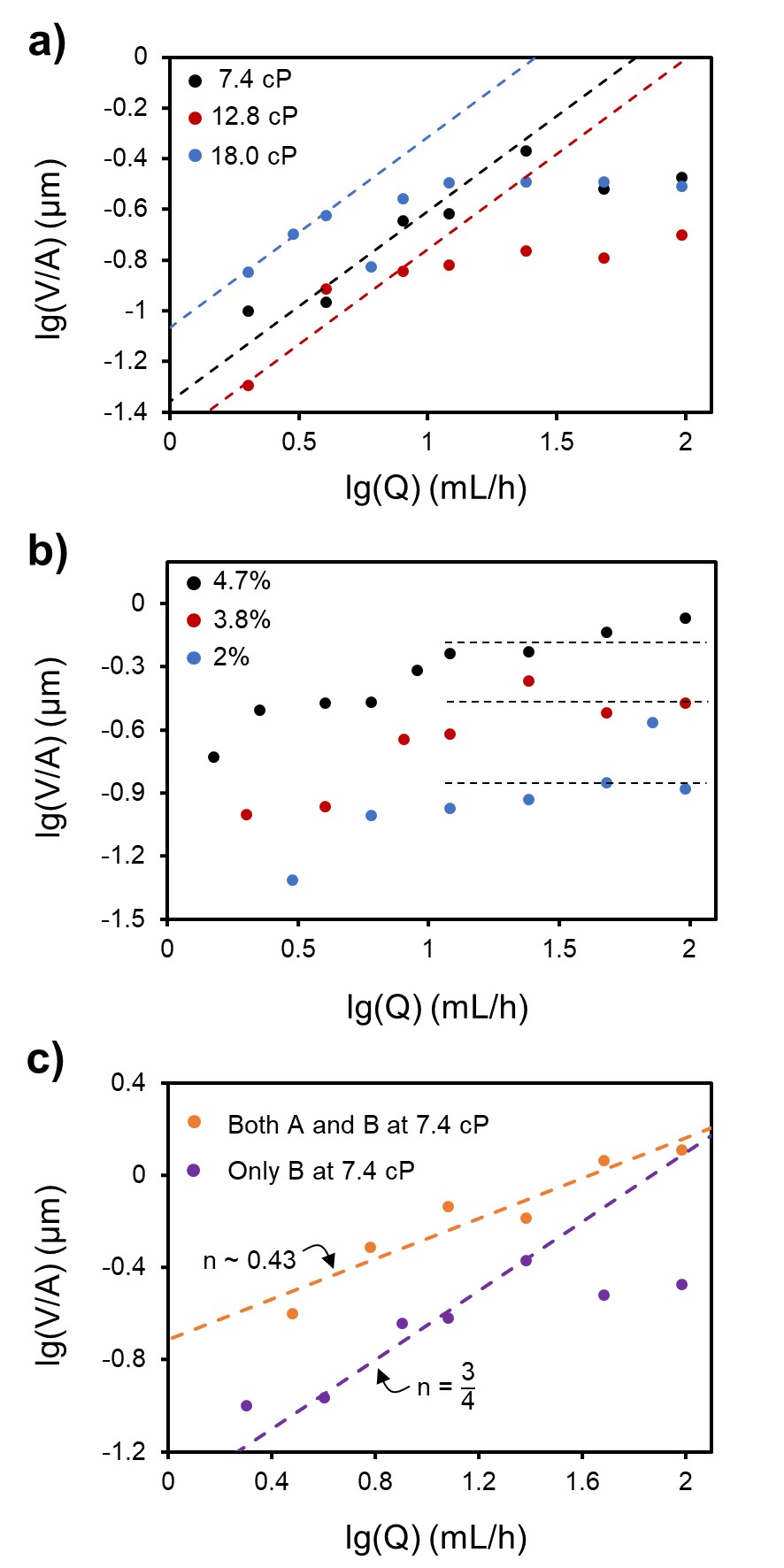}
\captionsetup{font={small}}
\caption{a) Droplet volume per unit area (V/A) produced using different flow rates (Q) at Solution B viscosity of 7.4, 12.8 and 18.0 cP. The dashed lines show the slope of 3/4 according to the scaling law $Vol_{f} \sim Pe^\frac{3}{4}$. For all viscosity values, the trend deviates at high flow rates. b) V/A produced using different Q at Solution B viscosity of 7.4 cP with different oil concentrations in Solution A. As shown by the dashed lines, V/A plateaus at high Q for all oil concentrations. c) V/A produced using different flow rates at viscosity of 7.4 cP for both Solution A and B (orange) or for Solution B only (purple). Orange dashed line in shows that while V/A scales with Q, the slope is lower than 3/4 (n $\sim$ 0.43). Purple dashed line with slope 3/4 shows the deviation of data at high Q.}
\label{plateau}
\end{figure}
\begin{figure}[ht]
\centering
\includegraphics[scale=1]{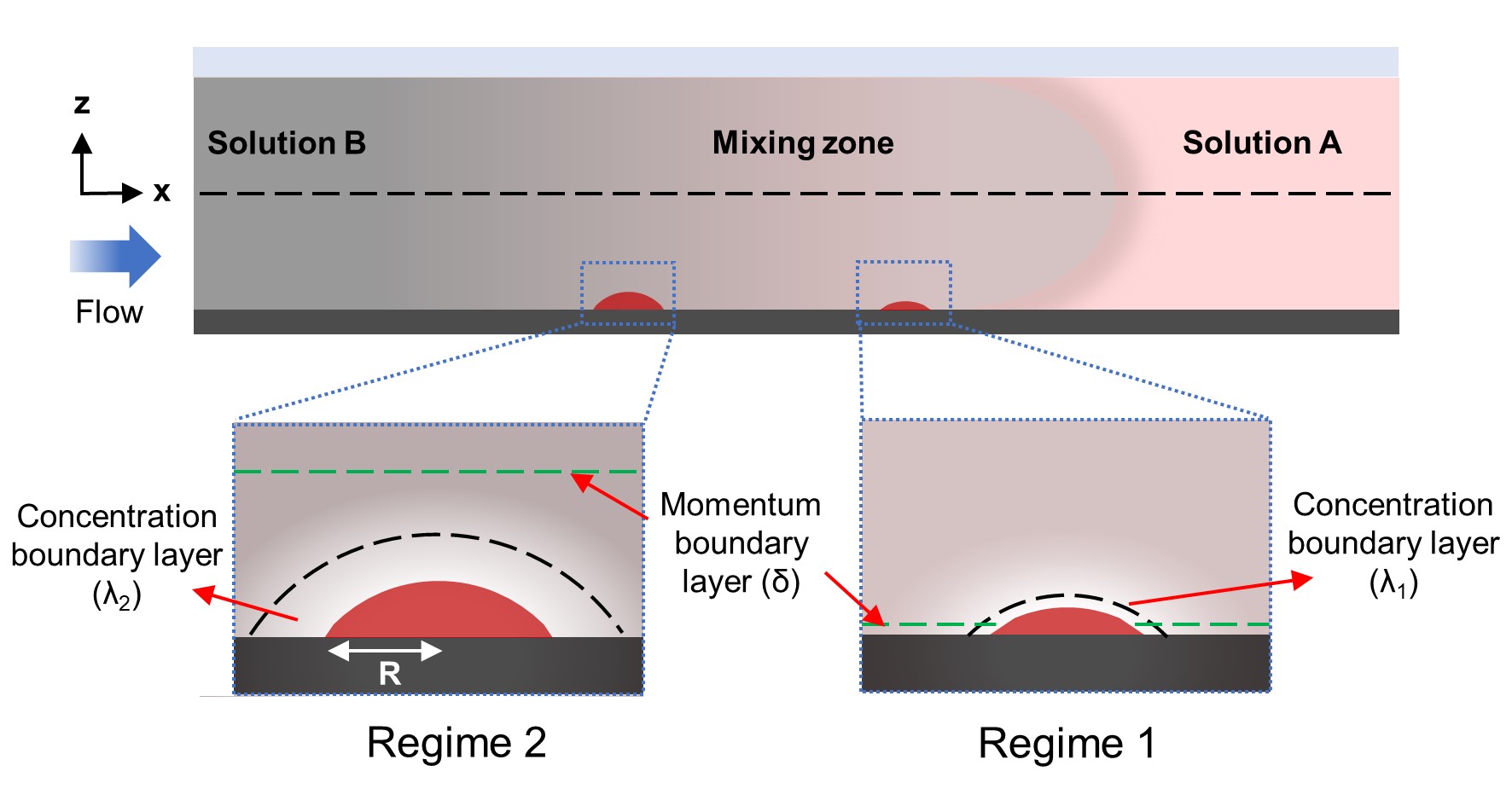}  
\captionsetup{font={small}}
\caption{Sketch showing two-regime growth model in a viscous solution. In regime 1, the concentration boundary layer is much thinner than the droplet and growth follows the scaling law. In regime 2, as viscosity increases, the concentration boundary layer becomes nested in the momentum boundary layer. At the same time, the concentration boundary layer becomes comparable to the size of droplet, minimizing the influence of Taylor dispersion and slowing down the growth.}
\label{regime}
\end{figure}

At low flow rates, the droplet growth in regime 1 is dominant while at fast flow rates, the growth in regime 2 prevails.  The critical flow rate where the transition from regime 1 to 2 occurs is influenced by the viscosity of Solution B.  At a higher viscosity of Solution B,  the transition from regime 1 to 2 takes place at a slower flow rate. Our analysis is consistent with the trend shown in Fig. \ref{plateau}. 

\indent Flow rate-independence is not observed when both viscosity of Solution A and B are increased to $\sim$ 7.4 cP (Fig. \ref{plateau}c). Droplet size increases with the flow rate of Solution B, but with a lower slope compared to the non-viscous case.  The final volume of droplets still scale with $Pe$ but with less dependence. The less dependence is from smaller droplets formed at higher flow rates, which may be attributed to reduced diffusivity of oil in a viscous medium. Less oil is transported from the mixing front to the growing droplets on the surface due to high viscosity in the time window before all the mixture is advected by the bulk flow.


\subsection{Influence of two-regime growth on droplet size distribution}
\begin{figure}[ht]
\centering
\includegraphics[scale=1]{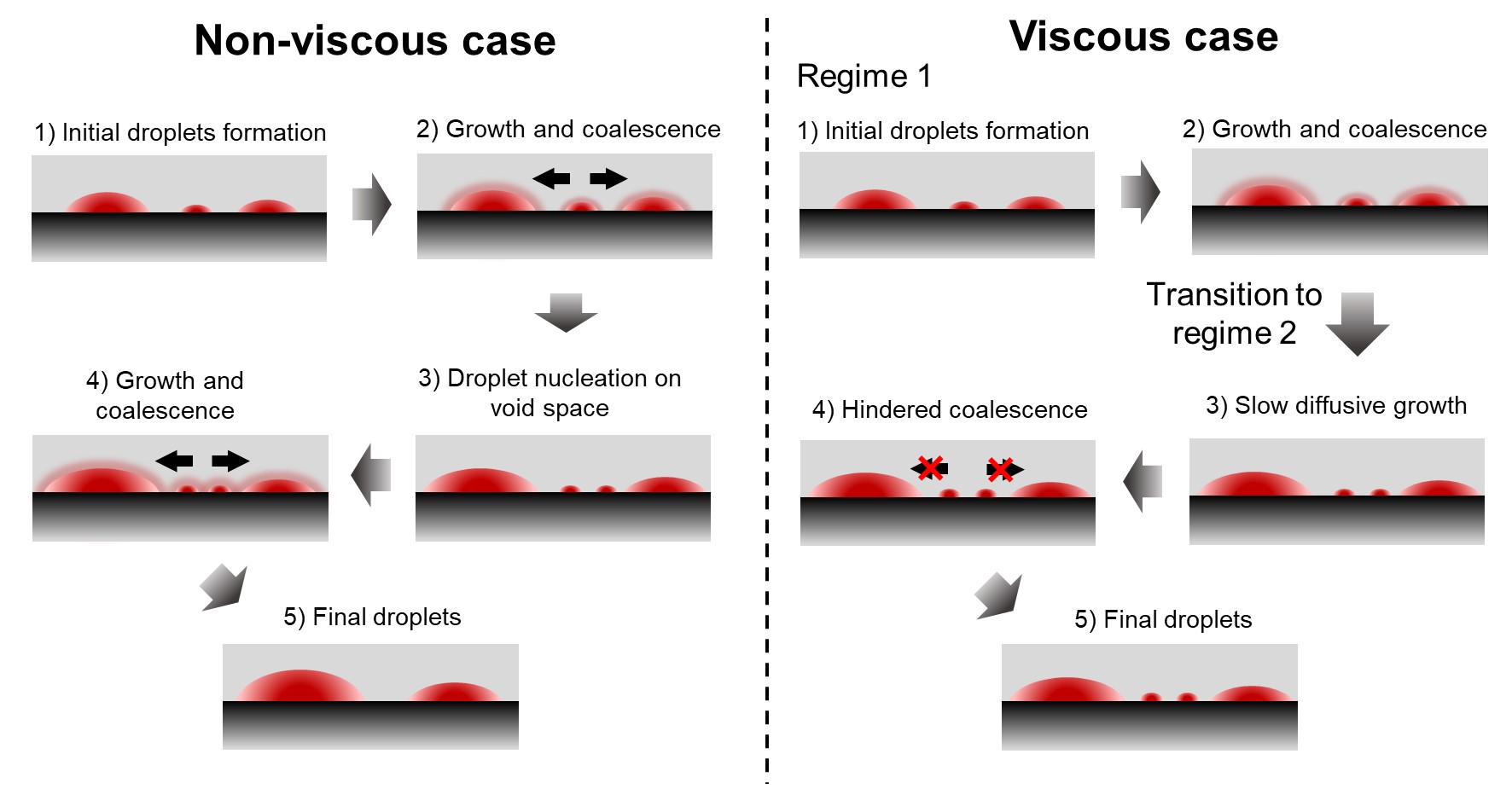}  
\captionsetup{font={small}}
\caption{Sketch showing the droplet growth modes in non-viscous case (left) and viscous case (right). In the non-viscous case, droplets form, grow and coalesce throughout the entire solvent exchange process. In contrast, when viscosity of Solution B is high, droplet only grows in regime 1 after which growth becomes slow and coalescence is inhibited due to high viscosity.}
\label{sketch}
\end{figure}

The two-regime growth mode may also explain the bimodal distribution of the droplet size. As Solution B is introduced and oversaturation occurs, the first generation of droplets in response to the oversaturation nucleate and grow till the surface coverage is close to a maximum of $\sim$ 50 $\%$.\cite{xu2017collective} 
This plateau is determined by the contact angle of the droplets on the surface \cite{xu2017collective}. Then the second generation of the droplets start to form on the bare substrate freed from the coalescence of the larger droplets. However, in contrast to the non-viscous case, the second generation of droplets do not  coalesce due to the high viscosity of the surrounding environment as sketched in Fig. \ref{sketch}. These small droplets do not grow much as the growth enters the slow regime 2. As a result, although a second generation of droplets form on the void spaces, coalescence between these droplets as well as with the first generation droplets is hindered because displacing the liquid between the droplets is now more difficult. Therefore by the end of solvent exchange, the droplets from second generation constitute the group of the small droplets resulting in a bimodal distribution of sizes.





\subsection{Enhanced detection of fluorescence by small droplets}
\begin{figure}[ht]
\centering
\includegraphics[scale=1]{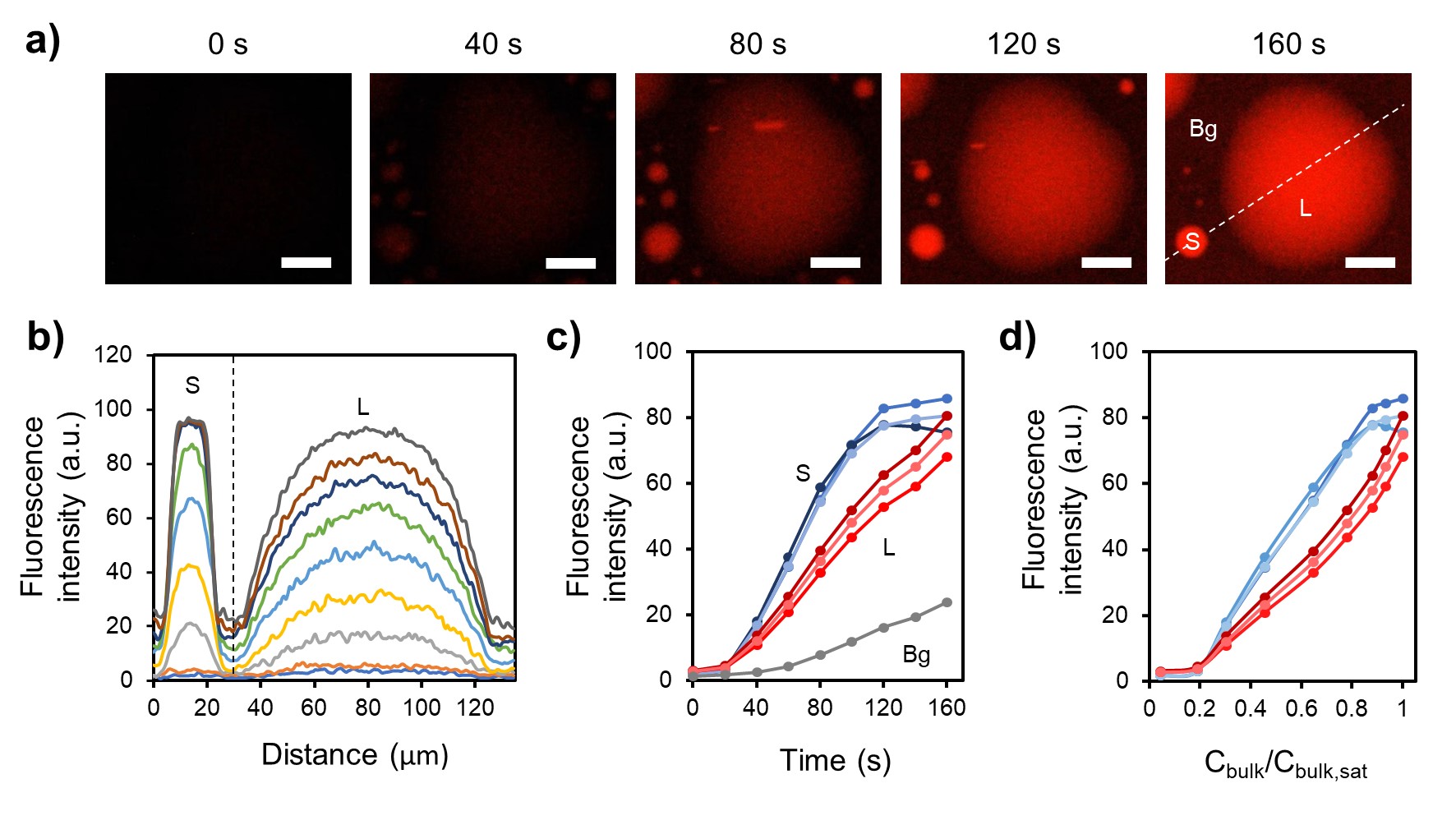}
\captionsetup{font={small}}
\caption{Enhancement of fluorescence intensity in small droplets a) Fluorescence images of two droplets at different times upon exposure to fluorescence dye solution. Scale bar $=$ 25 $\mu$m b) Intensity profile  through a straight line across the center of the two droplets as shown the last image of a) at different times. c) Intensity  of three small droplets (blue), three large droplets (red), and the background (grey) at different times. d) Intensity of small and large droplets for different concentrations of dye in the bulk relative to the saturation concentration in the bulk.}
\label{detection}
\end{figure}

In last section, we will demonstrate that the small droplets formed from viscous Solution B may be used for enhanced fluorescence detection. In our experiments water doped with a fluorescence dye (Nile Red) was injected into the chamber after the droplet formation in viscous Solution B. Once in contact with the solution, all droplets extract the dye from the solution. Our measurements show that the fluorescence intensity inside the droplets was significantly higher than the background as shown in Fig. \ref{detection}. Interestingly, the intensity increases faster inside small droplets compared to large droplets (Fig. \ref{detection}c).  
As the normalized concentration of the dye in the solution is around 80\% of the maximal concentration in the bulk flow, the intensity in the small droplets is about 0.7 $\sim$ 0.75, but only 0.45 $\sim$ 0.55 in the large droplets.

We attribute enhancement in the fluorescence intensity to lens effect of droplets. In previous reports, it has been shown that droplets can focus light to enhance the fluorescence intensity and enable more sensitive detection.\cite{zeng2016analchem,kaneko2019} The smaller droplets in our experiments have higher curvature so they can have stronger focusing effect for the emitted light from the fluorescence molecules. Such optical effect may find applications in chemical analysis, as fluorescence signal is widely used in sensing and biomedical analysis, for example the product from polymerase chain reaction in identification of nucleic acid sequence.

\section{Conclusion}
We study the effect of the viscosity of surrounding fluid on the droplet formation by solvent exchange. Different from the non-viscous case, the growth of droplets is independent of the flow rate when the viscous solution is fast.  We propose a two-regime growth model of droplets to explain the viscosity effect. In the first regime, droplet growth is still influenced by the external flow rate, while in the second regime the droplet growth relies on the diffusion of oil due to high viscosity.  The effect from the solution viscosity can be used to mediate final droplet sizes and to obtain distinct bi-model  distribution of the droplet size on a homogeneous surface, which may be applied in tuning the size distribution of droplets for various applications. As demonstrated in this work, strong intensity from fluorescence molecules is obtained from stable small droplets due to focusing effect.


\begin{acknowledgement}
This project was supported by the Natural Science and Engineering Research Council of Canada (NSERC) and Future Energy Systems (Canada First Research Excellence Fund). XZ acknowledges support from the Canada Research Chairs Program. The project was also partly supported by the ERC Proof-of-Concept grant. The authors thank Detlef Lohse for fruitful discussions.
\end{acknowledgement}

\begin{suppinfo}
\begin{itemize}
  \item Fig. S1: Schematic of solvent exchange process in a custom-built chamber and image of vertical setup for solvent exchange process to remove influence of gravity.
\end{itemize}

\end{suppinfo}

\bibliography{ref}

\newpage
\begin{table}[ht]
\captionsetup{font = {small}}
\caption{Experiment parameters for different Solution A composition (weight \%), Viscosity of Solution A viscosity and Solution B that is calculated based on the concentration of glycerol. \cite{cheng2008formula,volk2018density,Grunberg1949MixtureLF}}
\centering
\begin{tabular}{c|cccc}
\ & Solution A & Solution A & Solution B & Solution B \\
No.  & (ethanol:water: & viscosity & (water:glycerol) & viscosity \\
\ & HDODA:glycerol)& (cP)&  \  & (cP) \\
\hline
1&&&49: 51&$\sim$7.4\\
2&48.1: 48.1: 3.8: 0&$\sim$1&40: 60&$\sim$12.8\\
3&&&35: 65&$\sim$18.0\\
\hline
4&49: 49: 2: 0&&&\\
5&48.1: 48.1: 3.8: 0&$\sim$1&49: 51&$\sim$7.4\\
6&47.65: 47.65: 4.7: 0&&&\\
\hline
7&48.1: 11.5: 3.8: 36.6&$\sim$7.4&49: 51&$\sim$7.4\\
\end{tabular}
\end{table}
  
\setcounter{figure}{0}
\makeatletter 
\renewcommand{\thefigure}{S\arabic{figure}}

\end{document}